\documentclass[12pt]{iopart}

\usepackage{hyperref}
\hypersetup{colorlinks=true,linkcolor=blue,urlcolor=blue,citecolor=blue}
\usepackage{graphicx}
\usepackage[squaren, Gray, cdot]{SIunits}
\usepackage{amssymb}

\begin{document}

\title[]{Selective reflection from a Potassium atomic layer with a thickness as small as $\lambda /13$}

\author{A Sargsyan$^1$, E Klinger$^{1,2}$, C Leroy$^{2,4}$, I G Hughes$^3$, D Sarkisyan$^1$ \& C S Adams$^3$}

\address{$^1$ Institute for Physical Research -- National Academy of Sciences of Armenia, 0203 Ashtarak-2, Armenia}
\address{$^2$ Laboratoire Interdisciplinaire Carnot de Bourgogne -- UMR CNRS 6303, Universit\'e Bourgogne Franche-Comt\'e, Dijon, France}
\address{$^3$ Joint Quantum Centre (JQC) Durham-Newcastle, Department of Physics, Durham University, Durham, DH1 3LE United Kingdom}
\address{$^4$ Institute of Physics and Technology, National Research Tomsk Polytechnic University, Tomsk 634050, Russia}
\ead{emmanuel.klinger@u-bourgogne.fr}
\vspace{10pt}
\begin{indented}
\item[]
\end{indented}

\begin{abstract}
We demonstrate that a method using the derivative of the selective reflection signal from  a nanocell is a convenient and robust tool for atomic laser spectroscopy, achieving a nearly Doppler-free spectral resolution. The recorded linewidth of the signal from a potassium-filled cell, whose thickness $\ell$ lies in the range $350-500$~nm, is 18 times smaller than the Doppler linewidth ($\sim \unit{900}{\mega\hertz}$ full width at half maximum) of potassium atoms. We also show experimentally a sign oscillation of the reflected signal's derivative with a periodicity of $\lambda/2$ when $\ell$ varies from 190 to 1200~nm confirming the theoretical prediction. We report the first measurement of the van der Waals atom-surface interaction coefficient $C_3 = \unit{1.9\pm 0.3}{\kilo\hertz \times\micro\meter^3} $ of potassium $4S_{1/2} \rightarrow 4P_{3/2}$ transitions with the nanocell's sapphire windows, demonstrating the usefulness and convenience of the derivative of selective reflection technique for cell thicknesses in the range $60 -120~$nm.

\end{abstract}

%

\submitto{\JPB}
%
%
%

\section{Introduction}
Atomic vapors of alkali metals find a number of applications in quantum technologies such as chip-scale atomic clocks \cite{Knappe.APL.2004}, magnetoencephalography \cite{Sander.BOE.2012}, atomic vapor-based optical isolators \cite{Weller.OL.2012}, magnetic induction tomography \cite{Wickenbrock.OL.2014}, microwave imaging \cite{Horsley.NJP.2015},  narrow-band atomic filter \cite{Keaveney.OL.2018}, etc. For numerous applications, it is important to reduce the dimensions of the cell containing atomic vapor \cite{Kitching.APR.2018}, thus reaching the scale where the proximity of the atoms to the cell's surface becomes significant. Hence, a thorough understanding of the atom-surface interactions is essential. Many of the applications listed above use atoms in ground states, where atom-surface (AS) interactions are relatively small as the induced dipole is only a few Debye. However, AS interactions can still have a significant effect if the surface is located near the atom, that is less than typically 100~nm (see \cite{Bloch.AAMOP.2005} and refs. therein). In this regime, the AS potential is governed by an inverse power law 
$U_{\text{vdW}}=- C_\alpha z^{-\alpha}$, where $C_\alpha$ is the coupling coefficient and $z$ is the atom-surface distance. For an uncharged surface one expects the van der Waals (vdW) interaction with $\alpha =3$ in the regime $z<\lambda/2\pi \approx 120~$nm \cite{Haroche.PRL.1992,Peyrot.arxiv.2019}, where $\lambda$ is the transition wavelenght. Note that in the regime $z>\lambda/2\pi$, the AS interactation is descibed by the Casimir-Podler interaction with $\alpha=4$ \cite{Casimir.PR.1948,Laliotis.NC.2014}.

Recently, it was demonstrated that nanometric-thick optical cells (NCs) filled with alkali metal vapors are very convenient tools for alkali atom-surface studies using resonant absorption and fluorescence processes \cite{Bloch.AAMOP.2005,Peyrot.arxiv.2019,Fichet.EPL.2007,Keaveney.PRL.2012,Whittaker.PRL.2014,Whittaker.PRA.2015,Sargsyan.JPB.2016}. The NC is constructed such that its inner surfaces have a wedged form, thus the vapor column length $\ell$ can be varied smoothly in the range 50 -- 1500~nm with a simple vertical translation. Consequently, atoms inside the gap are located very close to the surface of the NC's dielectric windows (technical sapphire in our case).  The nanocell brings several benefits such as: (\textit{i}) a smooth translation of the laser beam in the vertical direction allows tunable distances atom-surface interactions studies; (\textit{ii}) when $\ell\sim 50~$nm, the efficiency of optical processes (e.g. absorption, fluorescence, selective reflection, etc.) is strongly reduced,  thus it is important to be able to increase the number density of the vapor which can be done by increasing the temperature of the NC. In our case, the temperature can be increased up to 500$~^\circ$C without any chemical reactions of hot alkali vapors with the sapphire windows of the NC. Note that in the case of glass cells, a strong chemical reaction at $\sim 150~^\circ$C causes a ``blackening'' of the cell's windows, making them opaque to the laser radiation.   

It is important to note that in order to obtain a detectable optical process signal when $\ell <100~$nm, the potassium vapor number density must be above $10^{13}~$cm$^{-3}$, which means that the temperature of the NC's reservoir must be above $150~^\circ$C, while the temperature at the windows must remain $20~^\circ$C higher in order to prevent condensation of the K vapor on the windows. That is why the NC (with sapphire windows) filled with Potassium vapor is ideal for atom-surface interaction study.

\section{Experiment}

\subsection{Design of the potassium-filled nanocell}
Figure.~\ref{fig:fig1}(b) shows a photograph of the potassium-filled NC. In order to have a variable thickness of atomic vapor column, a vertical wedged gap is formed using two \unit{1.5}{\micro\meter}-thick,  2~mm-long and 1~mm-wide  platinum strips, inserted between the NC's windows at the bottom (symmetrically from the both sides). It allows the gap thickness $\ell$ to vary from 50~nm (at the top) to 1500~nm (at the bottom). Note that platinum is resistant to highly corrosive hot alkali vapor.  The 2.3~mm-thick, $20 \times 30$~mm$^2$-large windows are made of well-polished crystalline sapphire. The c-axis is oriented perpendicular to the windows surface to reduce birefringence. Both windows of the NC are wedge-shaped for separation of the reflected beams. We have marked by an oval on the NC's picture (see Fig.~\ref{fig:fig1}(b)) the region in which the cell-gap thickness smoothly changes from 50~nm to 150~nm by vertical translation. The thin sidearm of the cell (reservoir) is filled with natural metallic potassium (93.23\% of $^{39}$K and 6.7\% of $^{41}$K). 

During the experiments, the reservoir (lower part of Fig.~\ref{fig:fig1}(b)) was heated to 150 -- 230$~^\circ$C to provide an atomic density of $N_\text{K} = 1\times 10^{13}$ to $5 \times 10^{14}$~ cm$^{-3}$. However, the potassium NC can successfully operate up to a temperature of $500~^\circ$C; its complete design is detailed in \cite{Whittaker.PRL.2014,Whittaker.PRA.2015}.  By changing the temperature of the reservoir in the range $50$ -- $500~^\circ$C, we can switch between the regimes $N_\text{K}k^{-3} \ll 1$ ($k=2\pi\lambda^{-1}$ is the incident light wave-vector), where dipole-dipole interactions are negligible, and $N_\text{K}k^{-3} \gg 100$, where dipole-dipole interactions dominate \cite{Weller.JPB.2011}.

\subsection{Experimental arrangement}
\begin{figure}[ht]
\centering
\includegraphics[scale=1]{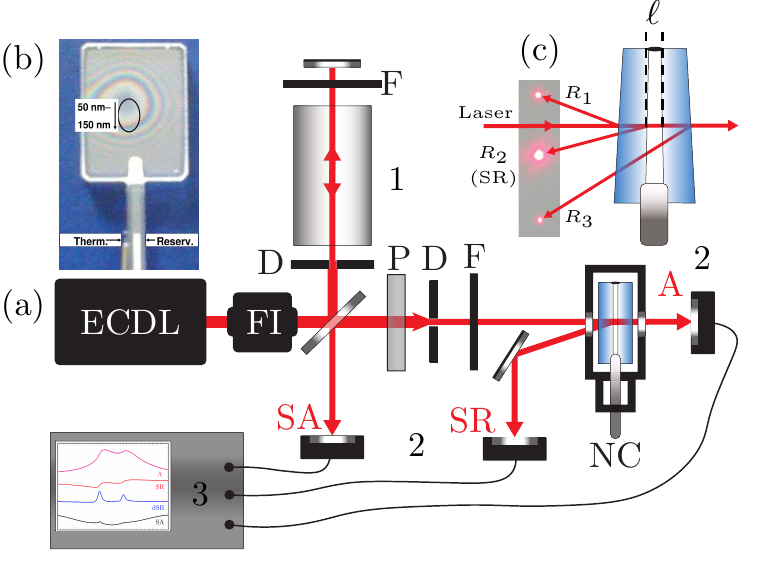}
\caption{(a) Layout of the experimental setup: ECDL -- cw tunable extended cavity diode laser, FI -- faraday isolator, NC -- potassium-filled nanocell in the oven,  1 -- frequency reference cm-long K cell, 2 -- photodetectors, 3 -- oscilloscope, P -- polariser, D -- diaphragm, F -- filter, SR -- selective reflection channel, SA -- saturated absorption channel, A -- absorption channel. (b) Photograph of the nanocell; the oval marks the region 50 -- 150~nm. (c) Geometry of three reflected beams from the NC; the selective reflection beam propagates in the direction of $R_2$.}
\label{fig:fig1}
\end{figure}

A schematic of the experimental setup is shown in Fig.~\ref{fig:fig1}(a). A continuous wave extended cavity diode laser with a wavelength $\lambda = 767$~nm,  a power of several mW and a spectral linewidth $\sim1~$MHz is used to scan across potassium D$_2$ line resonances. A Faraday isolator is used to prevent any laser radiation feedback into the cavity. The light polarisation is refined using a polariser, linearly-polarising the incident laser light. The laser beam diameter was lowered to 0.6~mm with the help of a diaphragm to shine the smallest area as possible on the cell while keeping an acceptable signal-to-noise ratio. 
Figure~\ref{fig:fig1}(c) shows the geometry of the three beams reflected from the NC, where SR is the beam selectively reflected from the interface between the windows and K atomic vapor; it propagates in the direction of $R_2$. For the selective reflection to be spectrally narrow, the laser radiation should be directed as close to normal incidence on the NC's window as possible \cite{Sargsyan.JCP.2016}. A fraction of laser radiation was guided to a reference unit forming a saturation absorption (SA) spectrum-based frequency reference from a 3~cm-long K cell \cite{Bloch.LP.1996}. The SR and SA radiation were recorded by sensitive photodiodes, outputting amplified signals later fed to a four-channel Tektronix oscilloscope. To separate the selective reflection signal from ambient noise, we used an interference filter ($\lambda = 767 $~nm) having a 10~nm bandwidth.

\begin{figure}[ht]
\centering
\includegraphics[scale=1]{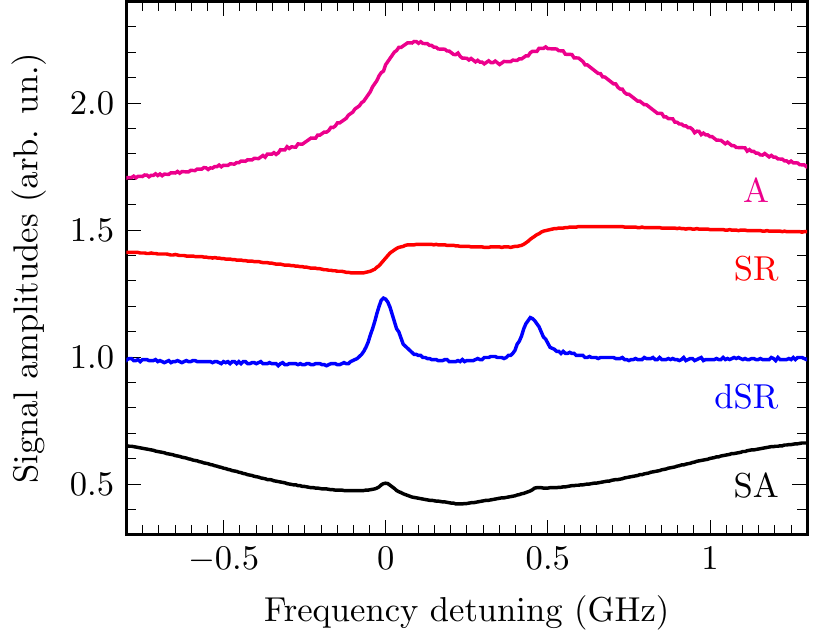}
\caption{$^{39}$K D$_2$ line spectra for the NC thickness $\ell =300$~nm,  the temperature $190~^\circ$C and the laser power 0.5~mW.  Magenta solid line (A) -- absorption spectrum; red solid line (SR) -- selective reflection spectrum (dispersive shape);  blue solid line (dSR) -- derivative of SR (absorptive shape). The linewidth of dSR is $\sim 60~$MHz (FWHM). The lower black solid line shows the reference SA spectrum. The curves have been shifted vertically for clarity.}
\label{fig:fig2}
\end{figure}

\section{Results and discussion}

The experimentally recorded spectra (absorption, selective reflection and derivation of selective reflection) of K D$_2$ line hyperfine transitions $F_g =1,2 \rightarrow F_e=0,1,2,3$ are shown in Fig.~\ref{fig:fig2} for a cell thickness $\ell=300$~nm. The temperature of the reservoir is 190$~^\circ$C, the laser power $P\sim 0.5$~mW. The upper magenta solid line curve shows the absorption spectrum. Although the Doppler linewidth of K transitions is more than 900 MHz, two groups of transitions corresponding to $F_g=1\rightarrow F_e=0,1,2$ and $F_g=2\rightarrow F_e=1,2,3$ separated by 462 MHz are partially resolved. The spectrum labeled SR (red solid line), having a dispersive shape, represents the selective reflection from the NC. The spectrum labeled dSR (blue solid line) corresponds to the derivative of the SR signal \cite{Sargsyan.JETPL.2016}. It is evident that the two groups of transitions of K D$_2$ line separated by 462 MHz are completely resolved in dSR spectrum, whose linewidth is $\sim 60$~MHz (FWHM) which is 15 times smaller than the Doppler linewidth of potassium atoms at $T_r = 190 ~^\circ$C. Note that, as the dSR provides a background-free signal, a laser power as small as several$\unit{}{\micro\watt}$ is enough if a sensitive photodetector is used. One should also note that the dSR signal amplitudes are proportional to the relative transition probabilities \cite{Sargsyan.JPB.2018}. The lower curve shows the SA used to provide a frequency reference.  

\begin{figure}[ht]
\centering
\includegraphics[scale=1]{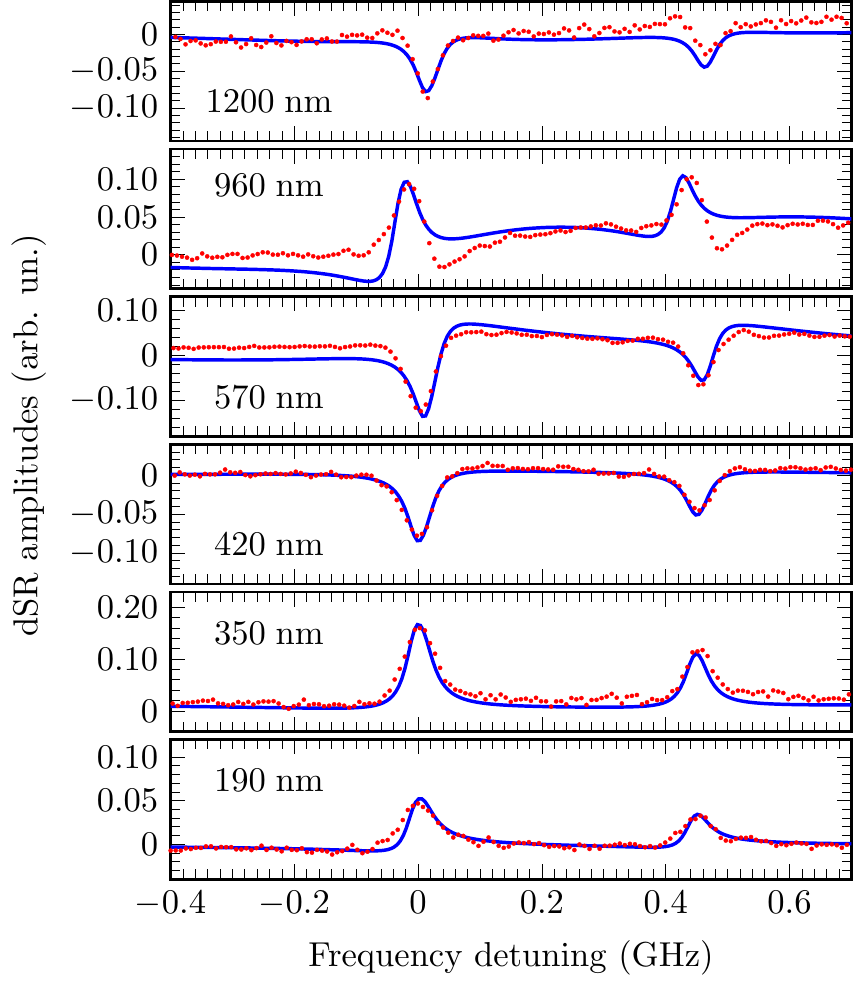}
\label{fig:fig3}
\caption{$^{39}$K D$_2$ line spectra as a function of the nanocell thickness. The experimental results for $T_r=190~^\circ$C, $P\sim0.2$~mW are shown with red dots; the corresponding theoretical curves are the blue solid lines. Oscillations of the dSR sign are observed: it is positive for $\ell=190,~350$ and $960$~nm, while negative for $\ell=420,~570$ and $1200$~nm. The theory coincides well with the experiment.}
\end{figure}

 In Fig.~\ref{fig:fig3}, we present the evolution of dSR spectra when the NC thickness is varied in the range 190 -- 1200~nm, for a reservoir temperature $T_r =190~^\circ$C and a laser power $P\sim0.2$~mW.  On this figure, we have superimposed the theoretical curves (blue solid lines) on the top of the experimental results (red dots). A similar situation to that obtained for Rb D$_1$ line \cite{Sargsyan.JETPL.2016} is observed: a small deviation from the thickness $\ell=\lambda/2= 384$~nm leads to a sign reversal of the dSR signal from negative for $\ell>\lambda/2$ to positive for $\ell<\lambda/2$. The same behaviour is also observed when the thickness varies from $\lambda/2< \ell<\lambda$ and again when $\lambda< \ell< 1.5 \lambda$. The predicted periodicity of the sign oscillation of $\lambda/2$ \cite{Vardanyan.PRA.1995, Dutier.JOSAB.2003} is thus verified experimentally over three periods. The small discrepancies seen between experiment and theory is due to the fact that the size of the homogeneous region (where $\ell$ is constant along the laser beam) gets smaller as the thickness is increased making the laser beam cover a larger range of thicknesses simultaneously. Note that for $\ell = m\lambda/2$, where $m$ is an integer, the SR (and thus dSR) signal vanishes due to destructive interference \cite{Dutier.JOSAB.2003}. 
 
 For applications in laser spectroscopy, the most convenient method is the dSR when $\ell \sim\lambda/2$, which correspond to the range 300 -- 450~nm for K D$_2$ studies. The linewidth of dSR presented in Fig.~\ref{fig:fig3} when $\ell=350$~nm is $\sim 50$~MHz (FWHM), which is slightly narrower than that  presented in Fig.~\ref{fig:fig2}, this difference is caused by the lower laser power. In the case of Fig.~\ref{fig:fig3} we achieve a spectral narrowing 18 times smaller than the potassium Doppler width. Note that in \cite{Sargsyan.JETP.2018} it was shown that under good experimental parameters ($\ell=350$~nm, $T_r=150~^\circ$C, $P\sim 0.1$~mW) the spectral linewidth of potassium D$_1$ line atomic transitions was reduced to 30~MHz using the dSR technique. 
 
\begin{figure}[ht]
\centering
\includegraphics[scale=1]{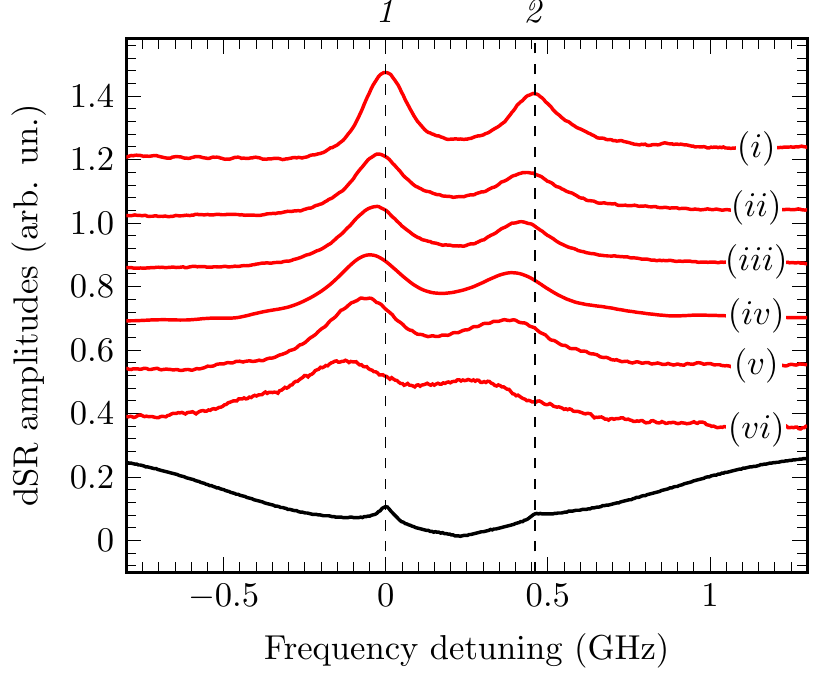}
\caption{Experimental $^{39}$K D$_2$ line spectra as a function of the nanocell thickness, for a temperature $T_r=230~^\circ$C and a laser power of 0.5 mW. ($i$) $\ell = 190 \pm 5$~nm: no frequency shift; ($ii$) $\ell = 120 \pm5$~nm: $-20$~MHz shift; ($iii$) $\ell =100 \pm 5$~nm: $-30$~MHz shift; ($iv$) $\ell = 90 \pm 5$~nm: $-42$~MHz shift; ($v$) $\ell=80 \pm5$~nm: $-60$~MHz shift and ($vi$) $\ell =60 \pm 5$~nm: $-130$~MHz shift. The lower black solid line curve serves as a frequency reference: \textit{1} corresponds to the group $F_g = 2\rightarrow F_e =1,2,3$ and \textit{2} corresponds to the group $F_g = 1\rightarrow F_e =0,1,2$ of $^{39}$K D$_2$ line.}
\label{fig:fig4}
\end{figure}

As visible in Fig.~\ref{fig:fig4}, a decrease in $\ell$ leads to a red frequency shift (\textit{i.e.} the frequency is decreasing, the wavelength is increasing) of the peak of the SR profile and to a strong asymmetric broadening toward low frequencies which is caused by vdW interactions of atoms with the NC's windows. As one can expect, atomic transition frequencies of the dSR spectrum for $\ell =350$~nm are unshifted when comparing with the SA frequency reference which is explained by a relatively large distance between atoms and NC's windows. Indeed, evidences of AS interactions for atomic D$_{1,2}$ lines of alkali metals are found only for thicknesses $\ell \leqslant 100$~nm, see \cite{Sargsyan.OL.2017}. For this reason, the atomic transition frequency shift is negligible for $\ell= 190$~nm since  atoms are still at a relatively large distance from the windows. Meanwhile, for $\ell= 120,~100,~90,~80$ and 60~nm, red frequency shifts of atomic transition peaks are $-20$, $-30$, $-42$, $-60$ and $-130$~MHz, respectively.  The thicknesses of the cell have been measured with a $\pm~5$~nm accuracy, using an interferometric method based on the Fabry-P\'erot nature of the NC, see \textit{e.g.} \cite{Sargsyan.JETP.2017}.

\begin{figure}[ht]
\centering
\includegraphics[scale=1]{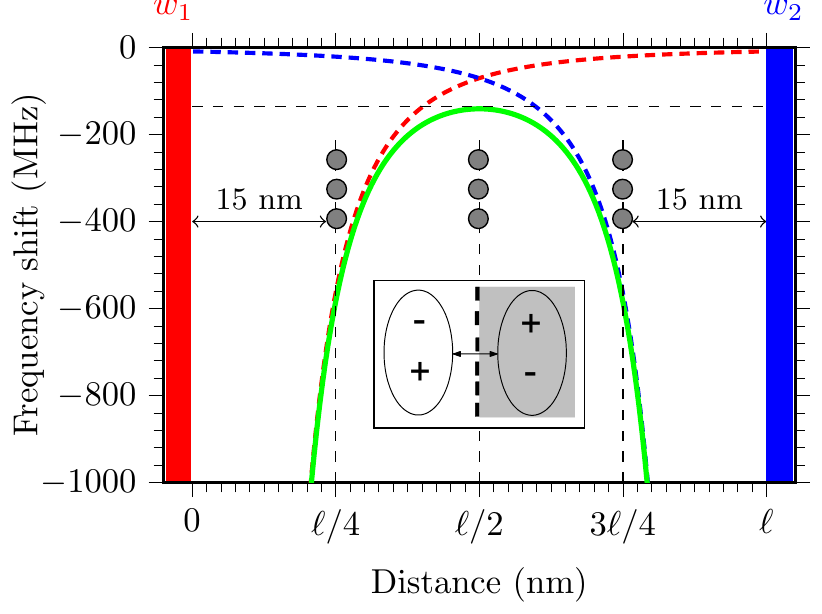}
\caption{Transition frequency shift versus atom-windows distance for the inter-window distance $\ell = 60$~nm. The red dashed line shows the red frequency shift caused by the first window ($w_1$) while the blue dashed one shows the one caused by the second window ($w_2$). The green solid line shows the total frequency shift caused by both windows simultaneously. The inset shows the sketch of an atom in the form of a dipole and its mirror image formed inside NC's dielectric windows. Potassium atoms located at the distance $z=15,~30,~45$~nm from the windows are represented as gray-filled circles.}
\label{fig:fig5}
\end{figure}

\begin{figure}[ht]
\centering
\includegraphics[scale=1]{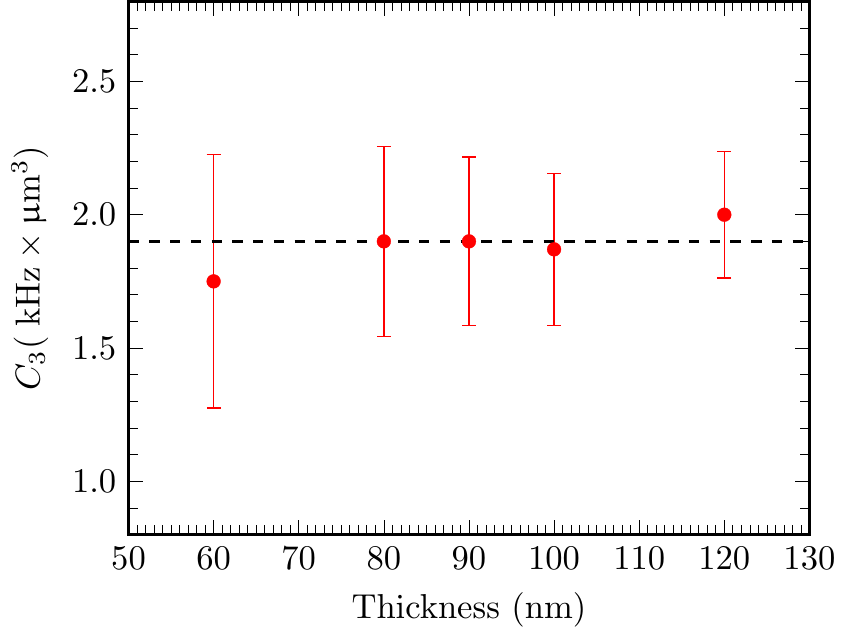}
\caption{ $^{39}$K D$_2$ line vdW interaction coefficient $C_3$ as a function of the NC thickness. The horizontal dashed line follows $C_3 = \unit{1.9}{\kilo\hertz\times\micro\meter^{3}}$.}
\label{fig:fig6}
\end{figure}

To determine the value of the vdW interaction coefficient $C_3$ of potassium $4S_{1/2} \rightarrow 4P_{3/2}$ transitions with nanocell's sapphire windows, we present in Fig.~\ref{fig:fig5} the frequency shifts with respect to the first window ($w_1$): $\Delta\nu_{\text{vdW}} = -C_3 z_1^{-3}$ (red dashed line), and for the second window ($w_2$): $\Delta\nu_{\text{vdW}} = -C_3 z_2^{-3}$ (blue dashed line), where $z_1$ and $z_2$ are the distances between a potassium atom and each windows ($w_1$) and ($w_2)$ respectively \cite{Bloch.AAMOP.2005}. The total frequency shift (green solid line) is the sum of both contributions, which can be expressed as
\begin{equation}
\Delta\nu_{\text{vdW}} = -\frac{C_3}{z^3} - \frac{C_3}{(\ell-z)^3}.
\label{eq:eq1}
\end{equation}
As is seen from Eq.~\ref{eq:eq1}, the total frequency shift caused by the influence of both windows for atoms located at the center of the NC, that is $z = \ell/2$, reads 
\begin{equation}
\Delta\nu_{\text{vdW}} = -16C_3\ell^{-3}. 
\label{eq:eq2}
\end{equation}
From Fig~\ref{fig:fig5}, one can see that the frequency shift of all atoms located at a distance $z=\ell/2 \pm5$~nm from the windows is $\sim|30|$~MHz close to the frequency shift at $z = \ell/2$ (for $\ell = 60$~nm). Meanwhile, the frequency shift of the atoms located at a distance $z=\ell/4 \pm 5$~nm (or $z=3\ell/4\pm 5$~nm) from the windows is larger by a few orders of magnitude (close to the frequency shift reached at $z = \ell/4$). Particularly, the red frequency shift experienced by atoms located at the distance $z=\ell/4 - 5$~nm (or $z=3\ell / 4 - 5$~nm) reaches $\sim3$~GHz. 
As a consequence, the spectral density of selective reflection signal is maximal for the atoms located near $z = \ell/2$ (center of the NC). Therefore the peak of the SR signal spectrum at any thickness $\ell\leqslant 100$~nm will be located where the  frequency shift modulus is minimal, see Fig.~\ref{fig:fig5}. Using the frequency shift values of the dSR signal peaks with $N_\text{K} \sim 4\times 10^{14}$~cm$^{-3}$ for various thicknesses (see Fig.~\ref{fig:fig4}) and Eq.~\ref{eq:eq2}, we have calculated the $C_3$ coefficient of the vdW interaction coefficient values, presented in Fig~\ref{fig:fig6}. Thus, we measure $C_3 = \unit{1.9 \pm 0.3}{\kilo\hertz\times\micro\meter^{3}}$ for the vdW interaction coefficient between $^{39}$K D$_2$ line transitions with NC's sapphire windows. Note that this value is close to that obtained for Cs D$_1$ line transitions \cite{Whittaker.PRA.2015}. The inaccuracy in the determination of the $C_3$ coefficient arises from the inaccuracy in determining the thickness of the nanocell.

 As already noted by Keaveney \textit{et al.} \cite{Keaveney.THESIS.2013}, for nanocell thicknesses $\ell > 200$~nm, the obtained spectrum broadens strongly for both D$_1$ and D$_2$ lines when increasing the temperature of the nanocell's reservoir because of atom-atom interaction, while not inducing any transition frequency shift. The situation for $\ell < 100$~nm is significantly different: as the temperature increases from 170 to 208$~^\circ$C, an additional red frequency shift to the vdW one has been observed, the cooperative Lamb shift. In order to measure a pure vdW interaction one has to minimize the influence of the atom-atom interaction; thus, it is preferable to measure the red shift of the SR signal peak at low densities, where the condition $N k^{-3}\ll 1$ is fulfilled. This condition remains fulfilled as long as $T_r\leqslant 160~^\circ$C (that is $N_\text{K} = 2\times 10^{13}$~cm$^{-3}$). However, a decrease in the temperature at 60 -- 90~nm strongly degrades the signal-to-noise ratio and, by extension, the accuracy of determining the position of the SR signal peak. At $160~^\circ$C, the red shift decreases by approximately 10\%.

Note that the behaviour of atomic transition frequencies when the potassium vapor density increases under the rigid spatial confinement by the windows of the nanocell is being actively studied, since new specific features have been revealed. For example, the following unusual behavior of the D$_2 $ line frequency of the atomic transition of potassium at a NC thickness $\ell = 490$~nm was revealed in a recent work \cite{Peyrot.PRL.2018} a blue frequency shift was detected when the K vapor density was increased such that $N_\text{K}k^{-3}\sim 1$; the frequency shift became zero with further increase of the number density (such that $N_\text{K}k^{-3}\sim 100$).

\section{Conclusion}

We have demonstrated that the use of derivative of the selective reflection method from nanocell containing potassium atomic vapor is a convenient tool for atomic laser spectroscopy, providing sub-Doppler spectral resolution of K D$_2$ line hyperfine transitions. The linewidth of dSR signal using a NC with the thickness $\ell=350$~nm is $\sim 50$~MHz (FWHM), which is 18 times smaller than the Doppler linewidth of K atoms at $T_r=190~^\circ$C. We demonstrated both experimentally and theoretically that when the cell thickness varies in the range 190 -- 1200~nm, sign oscillations of the dSR profile are observed with a periodicity of $\lambda/2$: when $ 0< \ell<\lambda/2$ the sign of dSR is positive; meanwhile, when $ \lambda/2< \ell<\lambda$ it changes to negative; finally, when $\lambda< \ell< 1.5 \lambda$ the sign changes back to positive. For applications in laser spectroscopy, the most convenient thickness for the dSR signal is $\ell\sim \lambda/2$. Let us also note another advantage of the dSR technique: in contrary to the well-known saturation absorption technique \cite{Smith.AJP.2004}, the dSR signal is proportional to the transition probabilities. We also show that the dSR method from a NC with a thickness in the range 60 -- 120 nm is a convenient tool for AS van der Waals interaction studies. Particularly, we were able to measure, for the first time,  $C_3 = \unit{1.9 \pm 0.3}{\kilo\hertz\times\micro\meter^{3}}$ for the vdW interaction coefficient between $^{39}$K D$_2$ line transition with NC's sapphire windows. Fabrication of NC whose windows are made of other dielectric materials resistant to hot aggressive alkali metal vapors will allows one to study the vdW interaction of atoms with another material and measure the corresponding $C_3$ coefficient.

The design of an all-glass vapor nanocell made from Borofloat glass is presented in \cite{Peyrot.OL.2019} and has been already implemented to study optical processes. A glass nanocell could make the SR technique accessible for many researchers.              

\ack
The authors acknowledge A Amiryan for assistance; J Keaveney and T Peyrot for fruitful discussions. A Sargsyan is thankful to the researchers from Durham University for the good scientific conditions provided during his visit. A Sargsyan and D Sarkisyan acknowledge support from the Committee on Science of the Ministry of Education and Science of the Republic of Armenia (project no. SCS 18T-1C018). E Klinger acknowledges a travel grant for collaborative research from FAST (Foundation for Armenian Science and Technology).



\section*{References}
\begin{thebibliography}{0}



\bibitem{Knappe.APL.2004} Knappe S, Shah V, Schwindt P D D, Hollberg L, Kitching J, Liew L-A and Moreland J 2004 \textit{Appl. Phys. Lett.} \textbf{85} 1460

\bibitem{Sander.BOE.2012} Sander T H, Preusser J, Mhaskar R, Kitching J, Trahms L, and Knappe S 2012 \textit{Biomed. Opt. Express} \textbf{3} 981

\bibitem{Weller.OL.2012} Weller L, Kleinbach K S, Zentile M A, Knappe S, Hughes I G, and Adams C S 2012 \textit{Opt. Lett.} \textbf{37} 3405

\bibitem{Wickenbrock.OL.2014} Wickenbrock A, Jurgilas S, Dow A, Marmugi L and Renzoni F 2014 \textit{Opt. Lett.} \textbf{39} 6367

\bibitem{Horsley.NJP.2015} Horsley A, Duand G-X and Treutlein Ph 2015\textit{ New J. Phys.} \textbf{17} 112002

\bibitem{Keaveney.OL.2018} Keaveney J, Wrathmall S A, Adams C S and  Hughes I G 2018 \textit{Opt. Lett.} \textbf{43}(17), 4272

\bibitem{Kitching.APR.2018} Kitching J 2018 \textit{Appl. Phys. Rev.} \textbf{5} 031302 

\bibitem{Bloch.AAMOP.2005} Bloch D and Ducloy M 2005 \textit{Adv. At. Mol. Opt. Phys.} \textbf{50} 91

\bibitem{Haroche.PRL.1992} Sandoghdar V, Sukenik C I, Hinds E A, and Haroche S 1992 Phys. Rev. Lett., \textbf{68}(23) 3432

\bibitem{Peyrot.arxiv.2019} Peyrot T, Šibalić N, Sortais Y R P, Browaeys A, Sargsyan A, Sarkisyan D, Hughes I G and Adams C S 2019 \textit{arXiv preprint} arXiv:1905.02783

\bibitem{Casimir.PR.1948} Casimir H B G and Polder D 1948 \textit{Phys. Rev.} \textbf{73} 360

\bibitem{Laliotis.NC.2014} Laliotis A, Passerat de Silans T, Maurin I, Ducloy M and Bloch D 2014 \textit{Nat. Commun.} \textbf{5} 4364

\bibitem{Fichet.EPL.2007} Fichet M, Dutier G, Yarovitsky A, Todorov P, Hamdi I, Maurin I, Saltiel S, Sarkisyan D, Gorza M P, Bloch D and Ducloy M 2007 \textit{Europhys. Lett.} \textbf{77} 54001

\bibitem{Keaveney.PRL.2012} Keaveney J, Sargsyan A, Krohn U, Hughes I G, Sarkisyan D and Adams C S 2012 \textit{Phys. Rev. Lett.} \textbf{108} 173601

\bibitem{Whittaker.PRL.2014} Whittaker K A, Keaveney J, Hughes I G, Sargsyan A, Sarkisyan D and Adams CS 2014 \textit{Phys. Rev. Lett. }\textbf{112} 253201

\bibitem{Whittaker.PRA.2015} Whittaker K A, Keaveney J, Hughes I G, Sargsyan A, Sarkisyan D and Adams CS 2015 \textit{Phys. Rev. A} \textbf{92} 052706

\bibitem{Sargsyan.JPB.2016} Sargsyan A, Pashayan-Leroy Y,  Leroy C and  Sarkisyan D 2016 \textit{J. Phys. B.} \textbf{49} 075001 

\bibitem{Weller.JPB.2011} Weller L, Bettles R J, Siddons P, Adams C S, and Hughes I G 2011 \textit{J. Phys. B.} \textbf{44}(19) 195006

\bibitem{Sargsyan.JCP.2016} Sargsyan A, Hakhumyan G, Sarkisyan A, Amiryan A and  Sarkisyan D 2016 \textit{J. of Contemp. Phys. (Armenian Ac. of Sci.)} \textbf{51}(4) 312

\bibitem{Bloch.LP.1996} Bloch D, Ducloy M, Senkov N, Velichansky V and Yudin V 1996  \textit{Laser Phys.} \textbf{6} 670 

\bibitem{Sargsyan.JETPL.2016} Sargsyan A, Klinger E, Pashayan-Leroy Y, Leroy C, Papoyan A and Sarkisyan D 2016 \textit{J. Exp. Theor. Phys. Lett.} \textbf{104} 224.

\bibitem{Sargsyan.JPB.2018} Sargsyan A, Klinger E, Tonoyan A, Leroy C and Sarkisyan D. 2018 \textit{J. Phys. B.} \textbf{51} 145001

\bibitem{Vardanyan.PRA.1995} Vartanyan T A and Lin D L 1995 Phys. Rev. A \textbf{51}(3) 1959

\bibitem{Dutier.JOSAB.2003} Dutier G, Saltiel S, Bloch D and Ducloy M 2003 \textit{J. Opt. Soc. Am. B} \textbf{20}(5) 793

\bibitem{Sargsyan.JETP.2018} Sargsyan A, Tonoyan A, Keaveney J, Hughes I G, Adams C S and Sarkisyan D. 2018 \textit{J. Exp. Theor. Phys.} \textbf{126} 293 


\bibitem{Sargsyan.OL.2017} Sargsyan A, Papoyan A, Hughes I G, Adams C S and Sarkisyan D 2017 \textit{Opt. Lett.} \textbf{42} 1476

\bibitem{Sargsyan.JETP.2017} Sargsyan A, Amiryan A, Cartaleva S and Sarkisyan D 2017 \textit{J. Exp. Theor. Phys.} \textbf{125}(1) 43

\bibitem{Keaveney.THESIS.2013} Keaveney J, Ph.D. thesis, Durham University, 2013, http://etheses.dur.ac.uk/7748/.

\bibitem{Peyrot.PRL.2018} Peyrot T, Sortais Y R P, Browaeys A, Sargsyan A, Sarkisyan D, Keaveney J, Hughes I G and Adams C S 2018 \textit{Phys. Rev. Lett.} \textbf{120} 243401

\bibitem{Smith.AJP.2004} Smith D A and Hughes I  G 2004 \textit{Am. Journ. Phys.} \textbf{72} 631

\bibitem{Peyrot.OL.2019} Peyrot T, Beurthe Ch, Coumar S, Roulliay M, Perronet K, Bonnay P, Adams C S, Browaeys A and Sortais Y R P  2019  \textit{Opt. Lett.} \textbf{44} 1940


\end{thebibliography}
\end{document}